\documentclass[useAMS,usenatbib,usegraphicx]{mn2e}

\usepackage{xcolor}
\usepackage{graphicx}
\usepackage{float}
\usepackage{color}
\usepackage{natbib}
\usepackage{lscape}
\usepackage{amssymb}
\usepackage{array}
\usepackage{ulem}

\title[Polarization in Puppis A]{New perspectives on the supernova remnant Puppis A based on a radio polarization study}

\author[Reynoso, Vel\'azquez, Cichowolski]{E. M. Reynoso$^{1}$\thanks{Email: ereynoso@iafe.uba.ar}, 
P. F. Vel\'azquez$^{2}$\thanks{Email: pablo@nucleares.unam.mx}, and S. Cichowolski$^{1}$\thanks{Email: scicho@iafe.uba.ar}\\
        $^{1}$Instituto de Astronom\'\i a y F\'\i sica del Espacio (IAFE), Av.
        Int. G\"uiraldes 2620, Pabell\'on IAFE, Ciudad Universitaria, \\Ciudad
        Aut\'onoma de Buenos Aires, Argentina\\
	$^{2}$Instituto de Ciencias Nucleares, Universidad Nacional Aut\'onoma de M\'exico,
	04510, Mexico City, M\'exico\\}

\begin{document}

   \date{Received 2018 January 29; accepted 2018 March 17}

\pagerange{\pageref{firstpage}--\pageref{lastpage}} \pubyear{2017}

\maketitle

\label{firstpage}
																		\begin{abstract}

We present a  polarization study towards the supernova remnant (SNR) Puppis A based on original observations performed with the Australia Telescope Compact Array (ATCA). Based on the analysis of a feature detected outside the SNR shell (called {\it `the tail'} throughout the paper), it was possible to disentangle the emission with origin in Puppis A itself from that coming from the foreground Vela SNR. We found a very low polarization fraction, of  about 3 percent on average. The upper limit of the magnetic field component parallel to the line of sight is estimated to be B$_\parallel \sim$ 20 $\mu$G. The statistical behavior of the magnetic vectors shows two preferential directions, almost perpendicular to each other, which are approximately aligned with the flat edges of Puppis A. A third, narrow peak oriented perpendicular to the Galactic plane suggests the existence of an interstellar magnetic field locally aligned in this direction. There is  evidence that the magnetic vectors along the shell are aligned with the shock front direction. The low polarization fraction and the statistical behavior of the magnetic vectors are compatible with a scenario where the SNR evolves inside a stellar wind bubble with a box-like morphology, produced by the interaction of the different stellar winds, one of them magnetized, launched by the SN progenitor. This scenario can furthermore explain the morphology of Puppis A, rendering little support to the previously accepted picture which involved strong density gradients to explain the flat, eastern edge of the shell.

\end{abstract}


\section{Introduction}\label{Int}

The vast majority of supernova remnants (SNRs) are strong emitters in radio 
wavelengths through the synchrotron mechanism, when relativistic electrons 
rotate around magnetic field lines. The synchrotron radiation is linearly 
polarized \citep{pach73}, hence polarimetric studies in the radio band of the electromagnetic spectrum constitute the main tool to investigate cosmic 
magnetic fields. Such studies provide information on the intensity of the
polarized emission, polarization degree and magnetic field orientation.

The obliquity of the SNR 
shock front with respect to the pre-shock magnetic field orientation may have 
consequences on the particle acceleration and magnetic field amplification processes \citep{rgb11}. Several models have been proposed to explain the geometry of SNRs on the basis of this obliquity \citep{fr90,BP16}. Radio polarization data towards SNRs sample shocked fields, while the direction of the interstellar (IS) pre-shock magnetic field remains unknown. However, from the analysis of the statistical distribution of the magnetic vectors, \citet{EMR+13} could infer that the IS field is parallel to the Galactic plane at the location of SN 1006, implying that the most efficient particle acceleration occurs when shocks are quasi-parallel to the magnetic fields.

The relatively young shell-type SNR Puppis A has a very peculiar morphology, with flat edges that imprint a somewhat boxlike shape to the shell. Infrared (IR) images at 24 $\mu$m and 70$\mu$m obtained with {\it Spitzer} \citep{IR2010} show bright filaments aligned parallel to the radio shell, also observed in X-rays with Chandra and XMM-Newton \citep{Xgd+13}. The bright, eastern side of the shell runs parallel to the Galactic plane. For decades, this flatness was ascribed to the collision of the shock front with a dense molecular cloud \citep[e.g.][]{gd+ma88,EMR+95}. However, unequivocal evidences of SNR-molecular cloud interaction, like line broadenings or OH masers at 1720 MHz \citep{frail+96}, have never been detected \citep[see also ][]{beate+00}. A recent H{\sc i} study \citep{pap2} places this SNR at a closer distance than previous estimates, where no external clouds or significant density gradients are observed. Numerical simulations \citep{af91} demonstrate that even under a density contrast as high as $10^3$, as would be needed to reproduce the flatness of the eastern edge of Puppis A, the resultant  morphology would be rather similar to the SNR VRO 42.05.01 (see their fig. 9), with a small shell towards the higher density region and a much larger shell on the opposite direction where the shock front progresses unimpeded. This is not observed in Puppis A, where the flat, eastern edge is followed by straight rims linked by sharp angled knees.

The straight outlines of the shell and filaments strongly suggest that, rather than interstellar medium (ISM) density gaps or gradients, magnetic fields could be playing a key role in the unique morphology of Puppis A. To explore this hypothesis, we analyze here new radio polarization observations carried out with the Australia Telescope Compact Array (ATCA). The angular extension of this SNR, about 50 arcmin in diameter, makes this source an ideal case for  a detailed study of the distribution of the magnetic fields. The first polarization study towards Puppis A was conducted by \citet{dm76}, who combined observations at 2.7 and 5 GHz performed with the 64-m radiotelescope at Parkes.  Very little 
polarization could be detected even at the brightest total intensity peaks, while unexplained discontinuities were found in the rotation measure (RM) maps. An improved polarization study was carried out almost 20 years later by \citet{MSH93} using observations at 8.4 and 4.75 GHz from the Parkes radiotelescope. They confirmed the very faint polarization reported by \citet{dm76}, with only 2-3 percent in most of the SNR and nearly 10 percent to the south, but found significant differences in the (RM) maps. 

This paper is organized as follows: the observations and data reduction to obtain the working images are described in Sect. \ref{Obs}. In Sect. \ref{Res} we explain the techniques applied to obtain the RM, magnetic vectors and depolarization and describe their distributions. These results are analyzed and discussed in Sect. \ref{Disc}. In Sect. \ref{Conc} we summarize our main conclusions.

\section{Observations and data reduction}\label{Obs}

Full spectropolarimetric observations of Puppis A in the radiocontinuum were 
obtained with the Compact Array Broadband Backend \citep[CABB;][]{CABB+11} of the ATCA in two 13 
hr runs during 2012, simultaneously with {\it `zoomed'} spectral observations 
in the H{\sc i} 21 cm and the four OH 18 cm lines. We covered the full extent of Puppis A by applying a mosaicking technique with 24 pointings arranged following the Nyquist theorem. The ATCA configurations were EW 352 and 750A for the first and second 13hr runs respectively, so that the visibilities correspond to baselines between 31 and 735 m excluding the 6th, fixed antenna. A thorough analysis of the 
continuum data was published in \citet{pap1}, while the H{\sc i} 21 cm line study was reported in \citet{pap2}. The primary flux and bandpass calibrator 
was PKS 1934-638, while PKS 0823-500 was used for phase and polarization 
calibration. The radio continuum was observed using a 2 GHz bandwidth divided 
into 2048 channels of 1 MHz width, centered at 1750 MHz. 

The data reduction was carried out with the {\sc miriad} software package 
\citep{Sault+95}. The observations in the EW 352 array were affected by a bug
which was fixed as described in \citet{pap2}. The {\it uv} data were split in eleven bands of 128 MHz, centered at frequencies from 1302 to 2582 MHz.  Data edition was performed for 
each band and linear feed (i.e. {\it XX, YY, XY, YX}) recorded with the CABB for source and calibrators, in order to remove interference and corrupted data. 

As the polarized radiation propagates parallel to the magnetic field in the
foreground ionized medium, the polarization plane rotates and therefore the measured direction of polarization is not coincident with the original one. This effect is known as Faraday rotation, and to accurately account for it, it is necessary to combine images at as many frequencies as possible to overcome the $n\pi$ ambiguity (see Sect. \ref{RM}). Hence, total intensity ($I$) images with the corresponding maps of the $Q$ and $U$ Stokes parameters were constructed for 10 different frequencies 
by merging visibilities from each two consecutive 128-MHz bands. A multifrequency
technique was used and the {\it uv}-range was optimized to retrieve the images
with a 100 arcsec beam. To perform a joint maximum-entropy deconvolution of the
mosaicked dirty images, we employed the miriad task {\sc pmosmem}, which can 
handle both negative and positive clean components found in the $Q$ and $U$ 
maps. To improve the signal-to-noise ratio, the final images were convolved 
with a circular 120-arcsec beam. In Table \ref{fqs} we display the 11 128-MHz 
bands in which the {\it uv} data were split indicating, for each of the 10 images constructed, their nominal frequencies and which band pairs were used.

\begin{table}
\begin{center}
\caption{Central column: central frequency of each 128-MHz band in which the 
{\it uv}-data were split. Left and right columns: index and frequency of each 
set of polarization images (Stokes $Q$ and $U$, total and polarized intensity, 
fractional polarization, and electric position angle), enclosing in the same 
raw the pair of bands (central column) that were combined to construct them.}
\begin{tabular}{cc|c|cc }
\hline
\multicolumn{2}{c|}{Polarization images}&Central frequency&
 \multicolumn{2}{|c}{Polarization images}\\
\cline{1-2} \cline{4-5}
\# &frequency&of {\it uv}-band &frequency&\#\\
&(MHz)&(MHz)&(MHz)\\
\hline
1&1383.200&1302\\
\cline{3-5}
&&1430&1437.283&2\\
\cline{1-3}
3&1651.866&1558\\
\cline{3-5}
&&1686&1755.213&4\\
\cline{1-3}
5&1872.070&1814\\
\cline{3-5}
&&1942&2018.097&6\\
\cline{1-3}
7&2112.696&2070\\
\cline{3-5}
&&2198&2265.315&8\\
\cline{1-3}
9&2392.916&2326\\
\cline{3-5}
&&2454&2524.846&10\\
\cline{1-3}
&&2582\\
\hline
\end{tabular}
\label{fqs}
\end{center}
\end{table}

\begin{figure}
\centering
\includegraphics[scale=0.45]{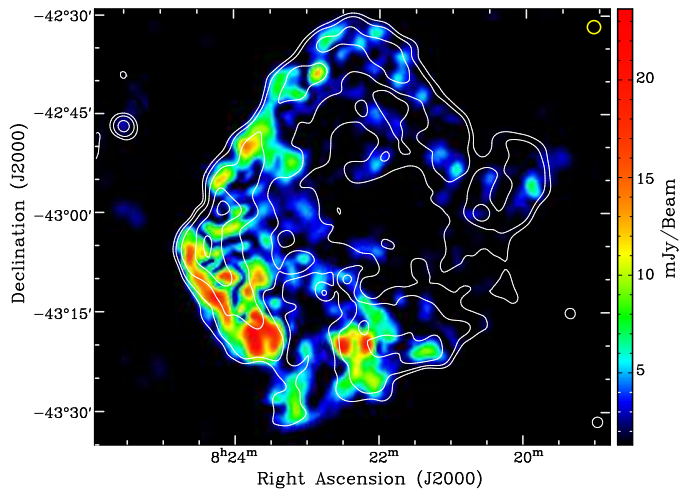}
   \caption{Polarized intensity at 2.24 GHz. The color scale, in mJy beam$^{-1}$, is shown in the bar at the right. Total intensity contours at 50, 100, 250, 600, and 1000 mJy beam$^{-1}$ are overlaid in white. The yellow circle at the top right corner represents the 2$^\prime$ beam.}
     \label{ipol}
\end{figure}

The Stokes $I$, $Q$, and $U$ images were combined with the miriad task {\sc 
impol} to obtain images of the polarized emission ($I_P$), polarization 
fraction ($\Pi = I_P/I$), and linear position angle\footnote{All angles throughout the paper are measured counter-clockwise from the north direction.} ($\chi$), given by
\begin{equation} \label{eq:xi}
\hskip 2 cm  \chi = {1\over 2} \tan^{-1} \big({U\over Q}\big), 
\end{equation}
together with their corresponding error images, which are computed simultaneously through the same task. The linear polarized emission is obtained from the Stokes $Q$ and $U$ parameters as $I_P = \sqrt{Q^2 + U^2}$, which results in positive values only. The Ricean bias,  consequence of this, was corrected to first order by subtracting the noise as $I_P = \sqrt{Q^2 + U^2 - \sigma^2}$, where $\sigma$ is the noise in the $Q$ and $U$ images, and varies for each frequency.  All pixels where $I_P$ was below 2$\sigma$ were blanked.

\begin{figure}
\centering
\includegraphics[scale=0.46]{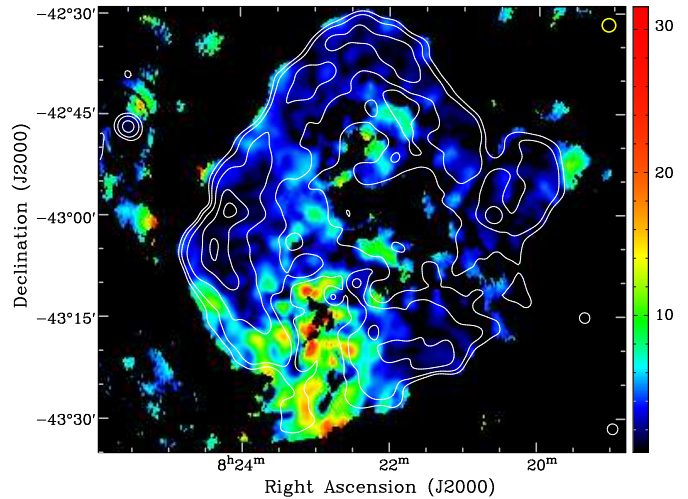}
   \caption{Fractional polarization distribution at 2.24 GHz. The color scale, 
in percentage of total intensity, is shown to the right. Total intensity contours at 50, 100, 250, 600, and 1000 mJy beam$^{-1}$ are overlaid in white. The yellow circle at the top right corner represents the 2$^\prime$ beam.}
     \label{fpol}
\end{figure}

Although the 10 images at different frequencies described above were useful to compute RM and polarization angle, as will be described in the next section, it is convenient to combine {\it uv}-data from a broader frequency range to obtain images with improved sensitivity. For that purpose, we used visibilities from 4 consecutive 128-MHz bands, from 2070 to 2454 MHz (see Table \ref{fqs}), to construct images of total and polarized emission, and combined them to compute polarization rate.  The error images were used to apply additional blanking to those pixels where the polarization rate error was either negative or higher than 0.05, and where the ratio between polarized intensity and its error was $\leq 3$. The resulting images with polarized emission and fractional polarization are shown in Figs. \ref{ipol} and \ref{fpol} respectively.

\section{Results}\label{Res}
\subsection{Rotation Measure}\label{RM}

\begin{figure}
\centering
\includegraphics[scale=0.46]{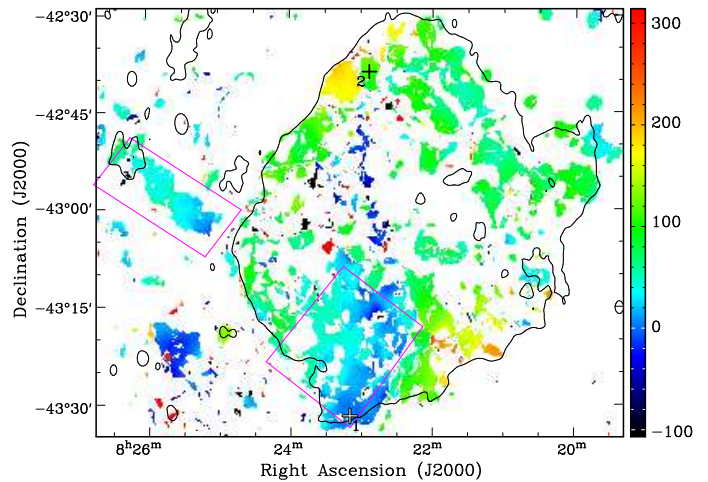}
   \caption{Rotation measure distribution towards Puppis A. The color scale is displayed in bar at the right, in units of rad m$^{-2}$. A few radio continuum black contours are overlaid. The beam size is 2$^\prime \times 2^\prime$. The plus signs numbered 1 and 2 are referred to the positions mentioned in Fig. \ref{rm-lines}. The magenta boxes enclose the two specific areas where the spatial distribution of the RM is discussed in the text (Sect. \ref{Disc}).}
     \label{rm-map}
\end{figure}

\begin{figure}
\centering
\includegraphics[scale=0.55]{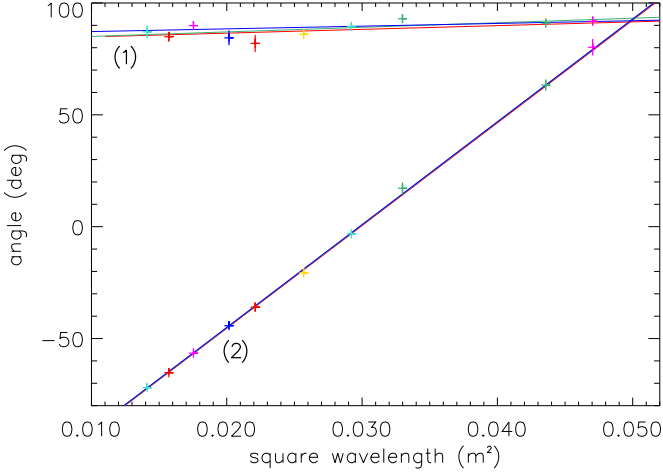}
 \caption{Linear fits of angle vs $\lambda^2$ towards two locations in 
Puppis A: (1) RA(2000)=8$^{\mathrm h}$23$^{\mathrm m}$11\hbox{$.\!\!{}^{\rm 
s}$}50, Dec.(2000)=$-43^\circ$31\hbox{$^{\prime}$}38\hbox{$.\!\!{}^{\prime\prime}$}5, 
and (2) RA(2000)=8$^{\mathrm h}$22$^{\mathrm m}$53\hbox{$.\!\!{}^{\rm s}$}41, 
Dec.(2000)=$-42^\circ$38\hbox{$^{\prime}$}38\hbox{$.\!\!{}^{\prime\prime}$}3. 
Each of the three fits performed for both locations was computed using 5 
points. To identify which set of 5 points was employed in each case, a same 
color code was used (red, green, or blue) for points and lines. When a point is 
used for more than one fit, the corresponding additive colors are used to plot 
it (cyan, yellow, or magenta). The average rotation measure for each position 
is (1) $\langle{\rm RM}\rangle=2.87\pm$0.65 rad m$^{-2}$ and (2) $\langle{\rm 
RM}\rangle=120.3\pm$0.2 rad m$^{-2}$.} 
\label{rm-lines}
\end{figure}

\begin{figure}
\centering
\includegraphics[scale=0.5]{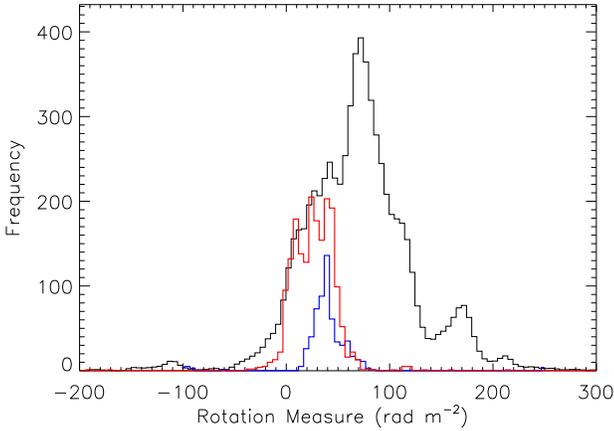}
   \caption{Statistical distribution of the rotation measure over Puppis A. The black histogram comprises the whole SNR, while the red and blue ones contain only the pixels within the boxes in the southern region and towards the east of the SNR, respectively, in Fig. \ref{rm-map}.}
     \label{rm-distr}
\end{figure}

\begin{figure*}
\centering
\includegraphics[scale=0.95]{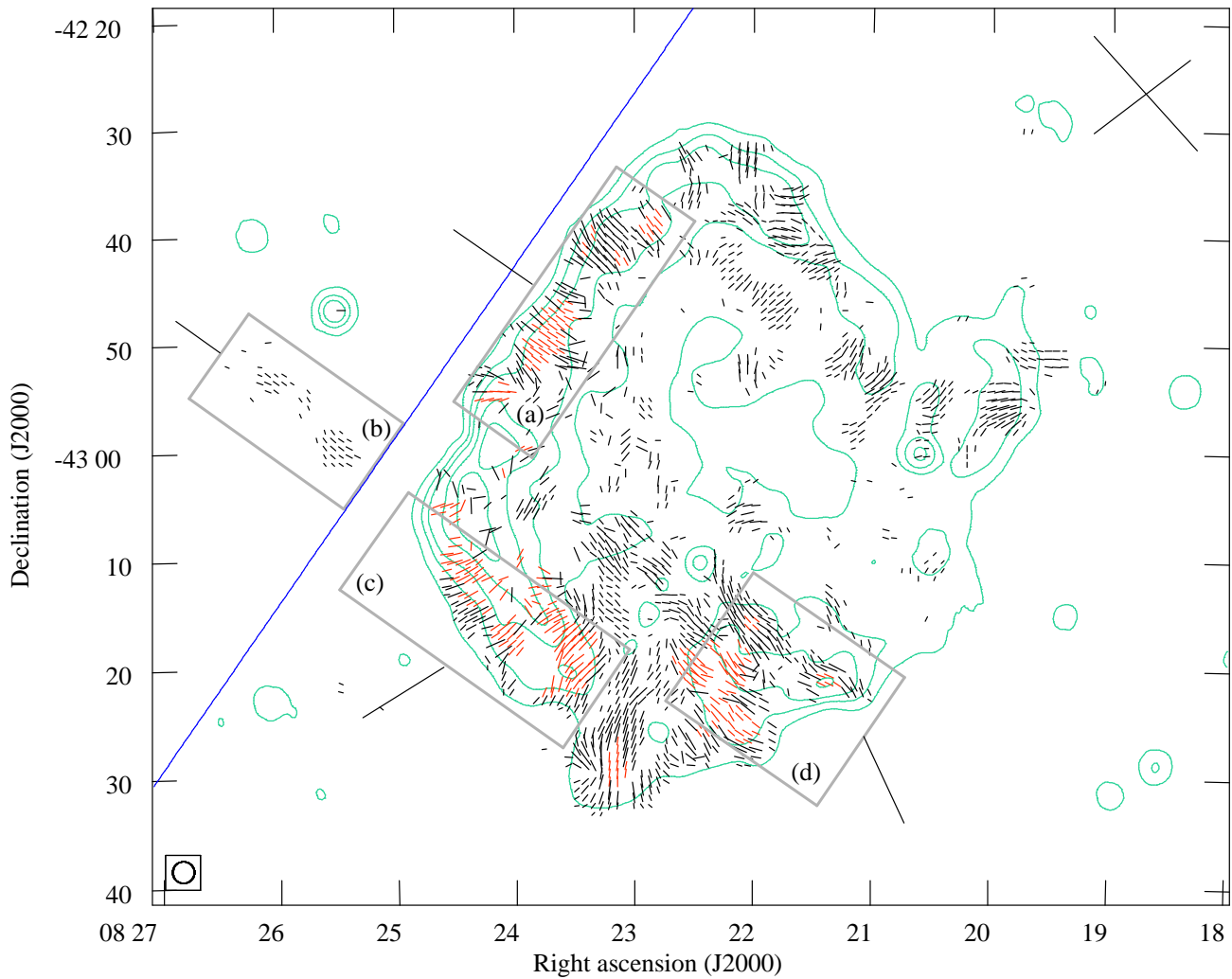}
   \caption{Distribution of magnetic field vectors on Puppis A corrected for Faraday rotation. One every 3 vectors is plotted in both axis. The overlaid green contours, at 30, 150, 300, 600, and 1000 mJy beam$^{-1}$, represent total intensity at 1.4 GHz. Pixels where the total intensity or the polarized flux are lower than 2.5 or 1.5 mJy beam$^{-1}$, respectively, are blanked. For clarity, different colors and scales have been used for the magnetic vectors: if the polarized flux (see Fig. \ref{ipol}) is above (below) a threshold of 7.5 mJy beam$^{-1}$, vectors are plotted in orange (black) and a length of 1 arcmin is equivalent to a polarized flux of 10 (4) mJy beam$^{-1}$.  To indicate the direction of the Galactic plane, a blue line at constant Galactic Latitude, $b=-3^\circ$, was added. The beam, 2$^\prime$, is plotted inside a box at the bottom, left corner. The regions enclosed by the 4 grey boxes labeled $a - d$ are discussed in Sect. \ref{Disc}. The black lines protruding from boxes {\it a} and {\it b} are orthogonal to the Galactic plane, while those drawn in boxes {\it c} and {\it d} approximately represent the normal to the corresponding flat rim. The large cross at the upper, right corner indicates the two main directions derived from Fig. \ref{vall}.}
     \label{vecs}
\end{figure*}

\begin{figure}
\centering
\includegraphics[scale=0.5]{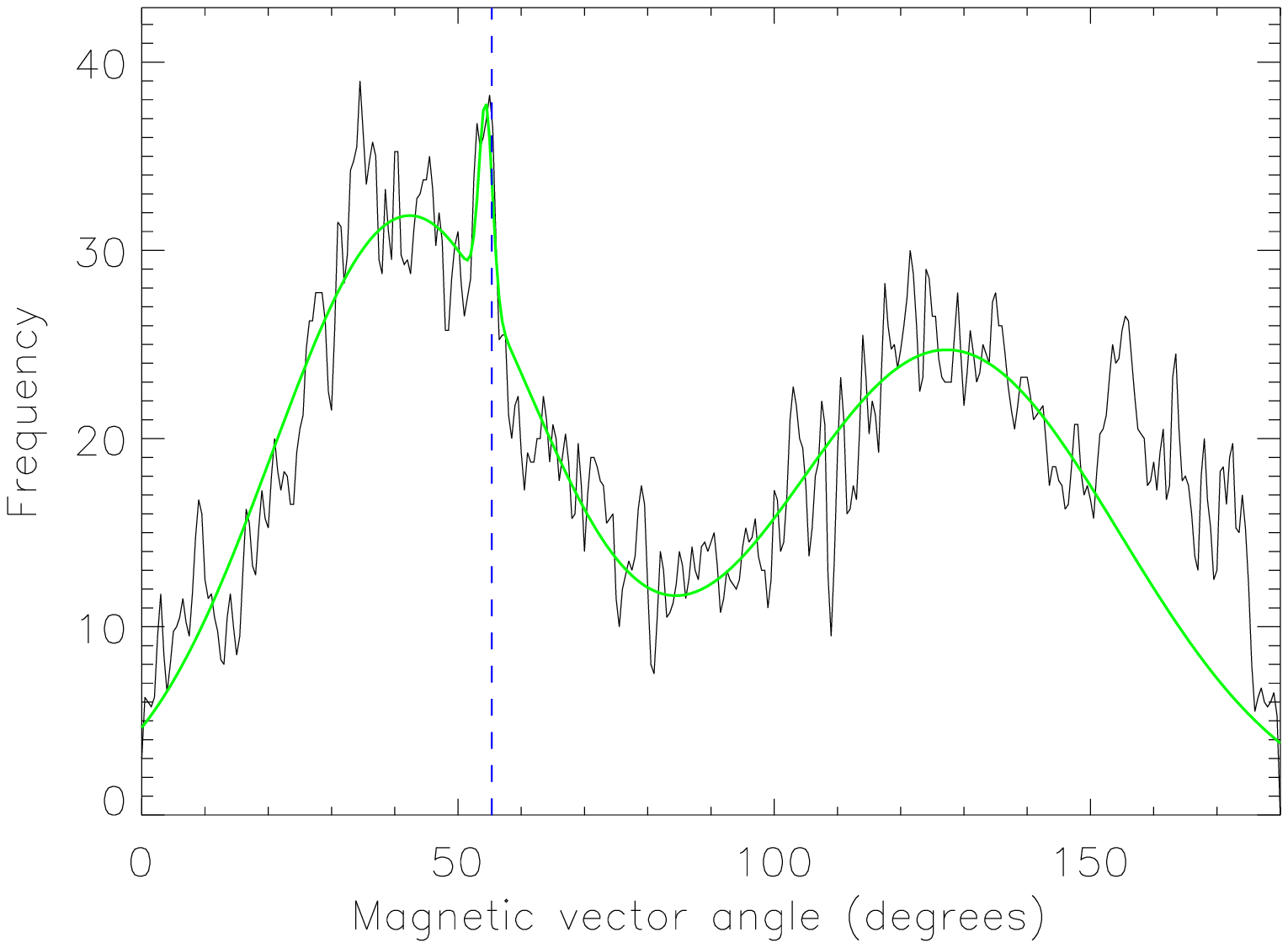}
   \caption{Distribution of the magnetic vectors angle over Puppis A measured counter-clockwise from the north. The fit is plotted with a green solid line. The vertical dashed line is the direction normal to the Galactic plane.}
     \label{vall}
\end{figure}

The linear position angle $\chi$ of the electric vectors determined by $Q$ and $U$ through equation (1) is the observed angle after the intrinsic angle $\chi_0$ undergoes a rotation as the linearly polarized light propagates through a foreground magnetized plasma. A correction 
must therefore be applied to retrieve $\chi_0$. The angular rotation of the
polarization plane $\Delta \chi$ is defined as
\begin{equation} \label{eq:lamd}
\hskip 2 cm \Delta \chi(\lambda) = \chi(\lambda) - \chi_0 = {\rm RM}\, \lambda^2,
\end{equation}
where $\chi$ and $\chi_0$ are expressed in radians and $\lambda$, in m. RM is given by
\begin{equation} \label{eq:rm} \hskip 1 cm
{\rm  RM}={{e^3}\over{\epsilon_0 8 \pi^2 m_e^2 c^3}} \int^r_{r_0} n_e \vec{B}.d\vec{r} 
= 0.812 n_e B_{\parallel} \Delta r
\end{equation} 
in units of rad m$^{-2}$, where $n_e$ is the number density of the thermal electrons (in units of $\mathrm{cm^{-3}}$), $B_{\parallel}$ is the magnetic field component in the direction of the line of sight (in $\mu$G) and $\Delta r$ is the depth between source and observer (in pc). For the expression at the right in Eq. (\ref{eq:rm}), a constant (or average) value of $n_e$ and $B_{\parallel}$ are considered.

The main problem with deriving RM is that there can be an extra $n\pi$ in 
$\Delta \chi$ when position angles at only two consecutive frequencies are 
considered. To overcome this obstacle, it is necessary to obtain position 
angles in as many frequencies as possible. The task {\sc imrm} implemented in 
{\sc miriad} allows to fit position angle images from up to five different 
frequencies on a pixel by pixel basis, removing the $n\pi$ ambiguity. In the 
same process, the task computes the unrotated angle $\chi_0$. 

To obtain a reliable RM, we ran {\sc imrm} on three different groups of five 
$\chi(\lambda) - \lambda^2$ pairs selected from the 10 sets of polarization 
images produced, listed in table \ref{fqs}. Linear fits of $\chi$ vs $\lambda^2$
were computed for each pixel, and the result with the corresponding error were
recorded in two separate maps, blanking those pixels for which the goodness of the fitting was less than 0.001. Finally, the three RM maps obtained were
averaged and an associated error image was computed. Pixels with errors $\geq$
5 rad m$^{-2}$, which amounted to only 5\% of the valid pixels, were blanked. The final image with the RM distribution is shown in Fig. \ref{rm-map}.

\begin{figure}
\centering
\includegraphics[scale=0.46]{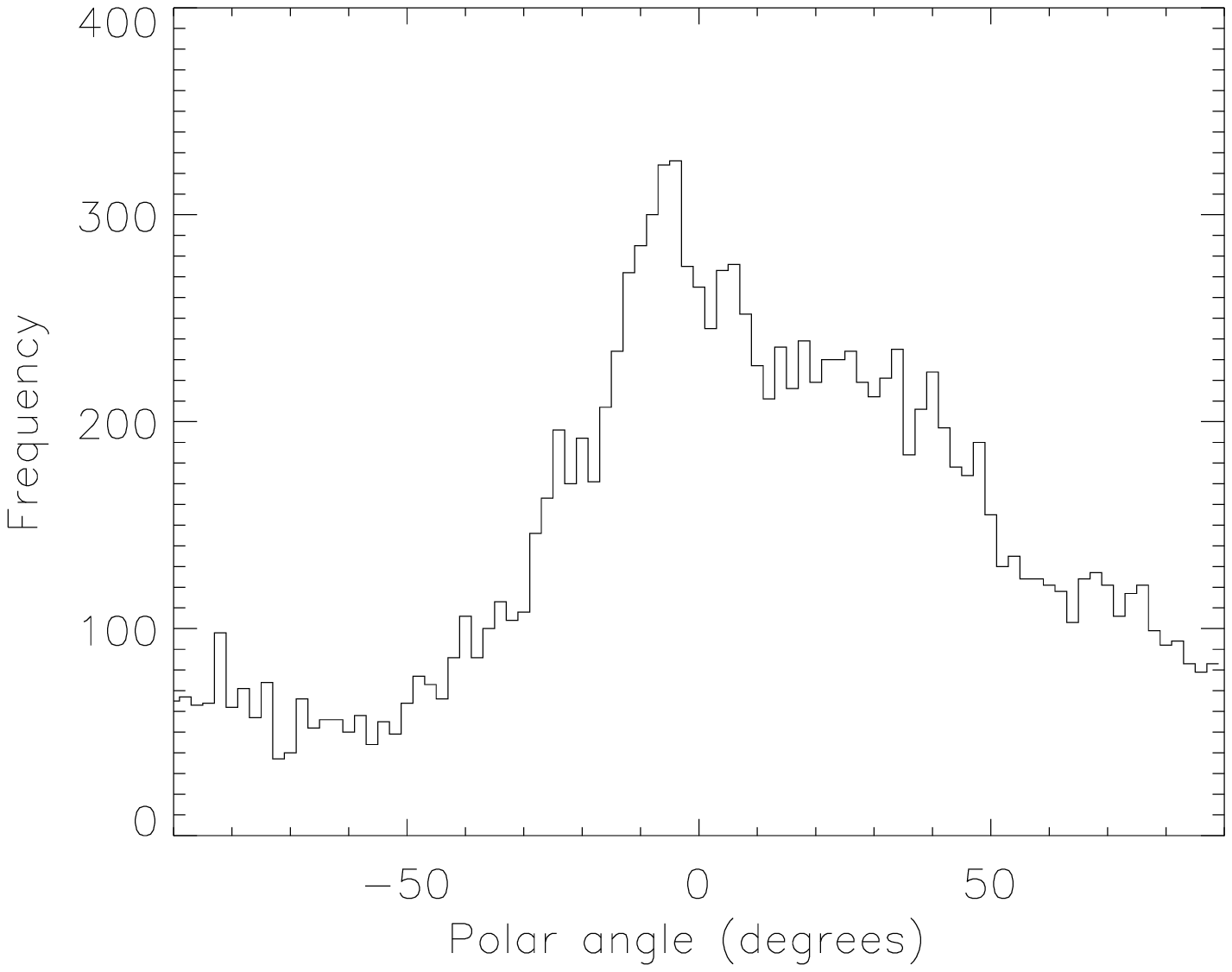}
   \caption{Distribution of the directions of the magnetic vectors over Puppis A based on a polar reference system. Radial vectors are at 0$^\circ$, while tangential vectors are at $\pm 90^\circ$.}
     \label{pola}
\end{figure}

\begin{figure*}
  \includegraphics[width=0.49\textwidth]{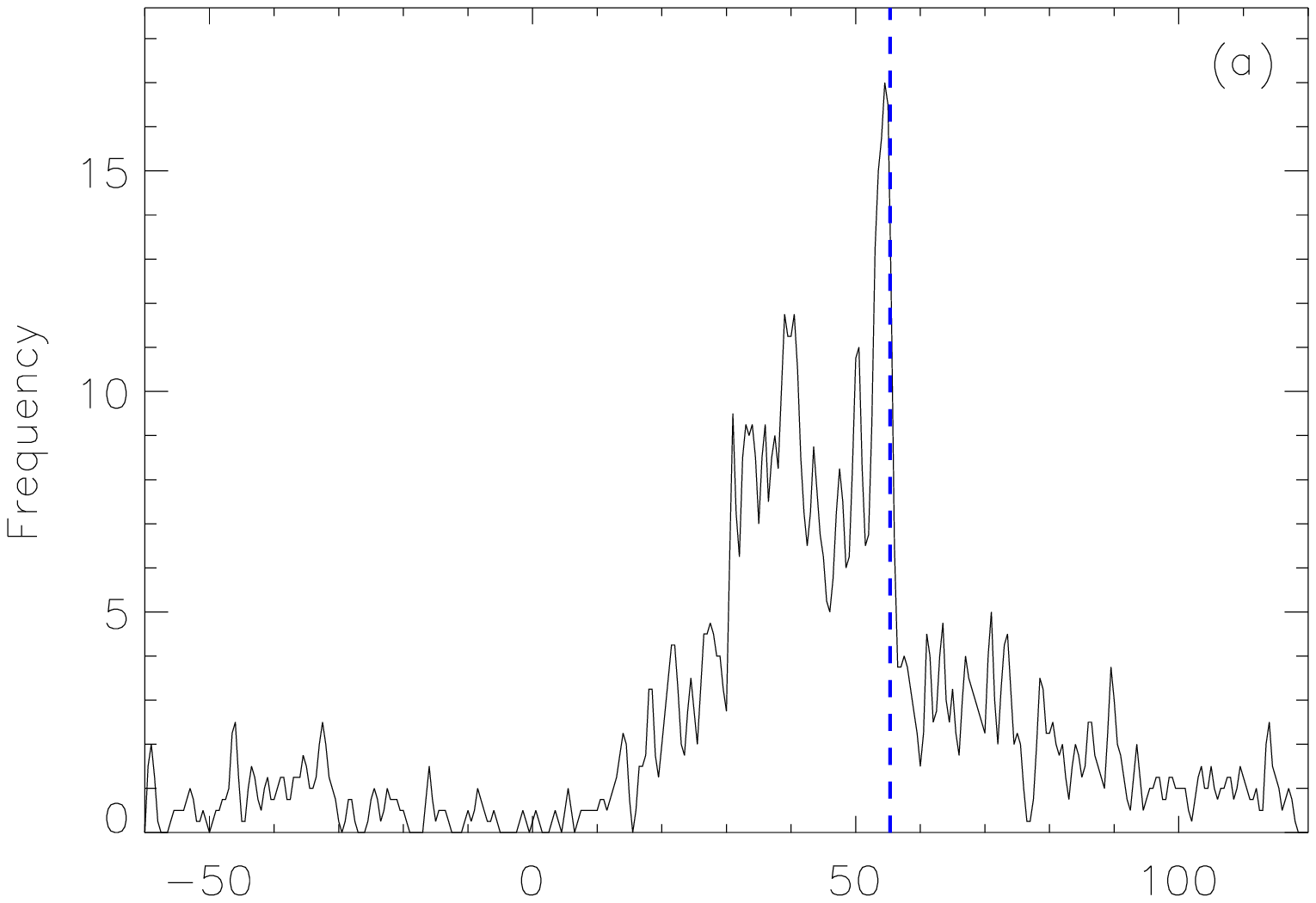}
    \hfill
  \includegraphics[width=0.475\textwidth]{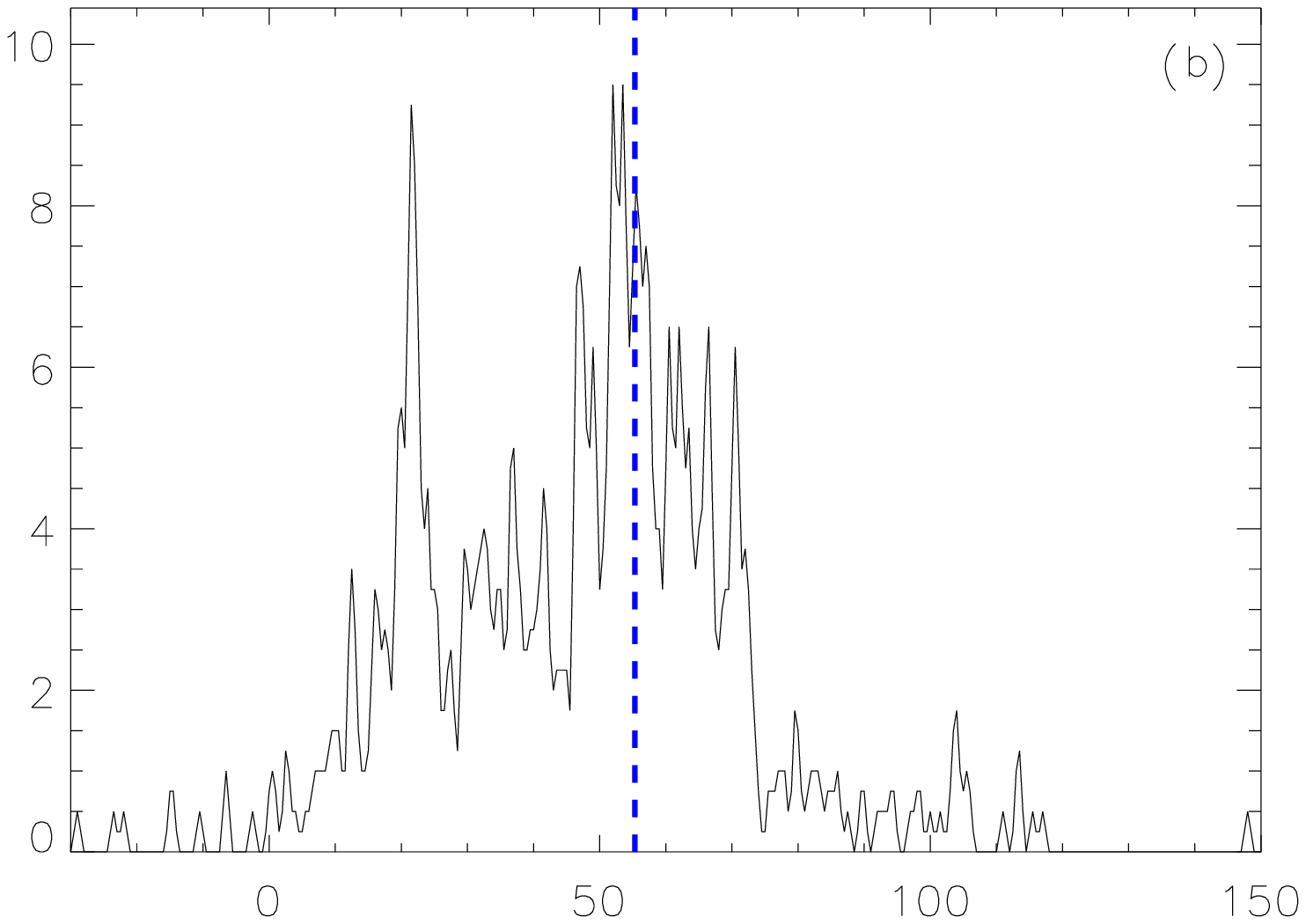}
  \includegraphics[width=0.49\textwidth]{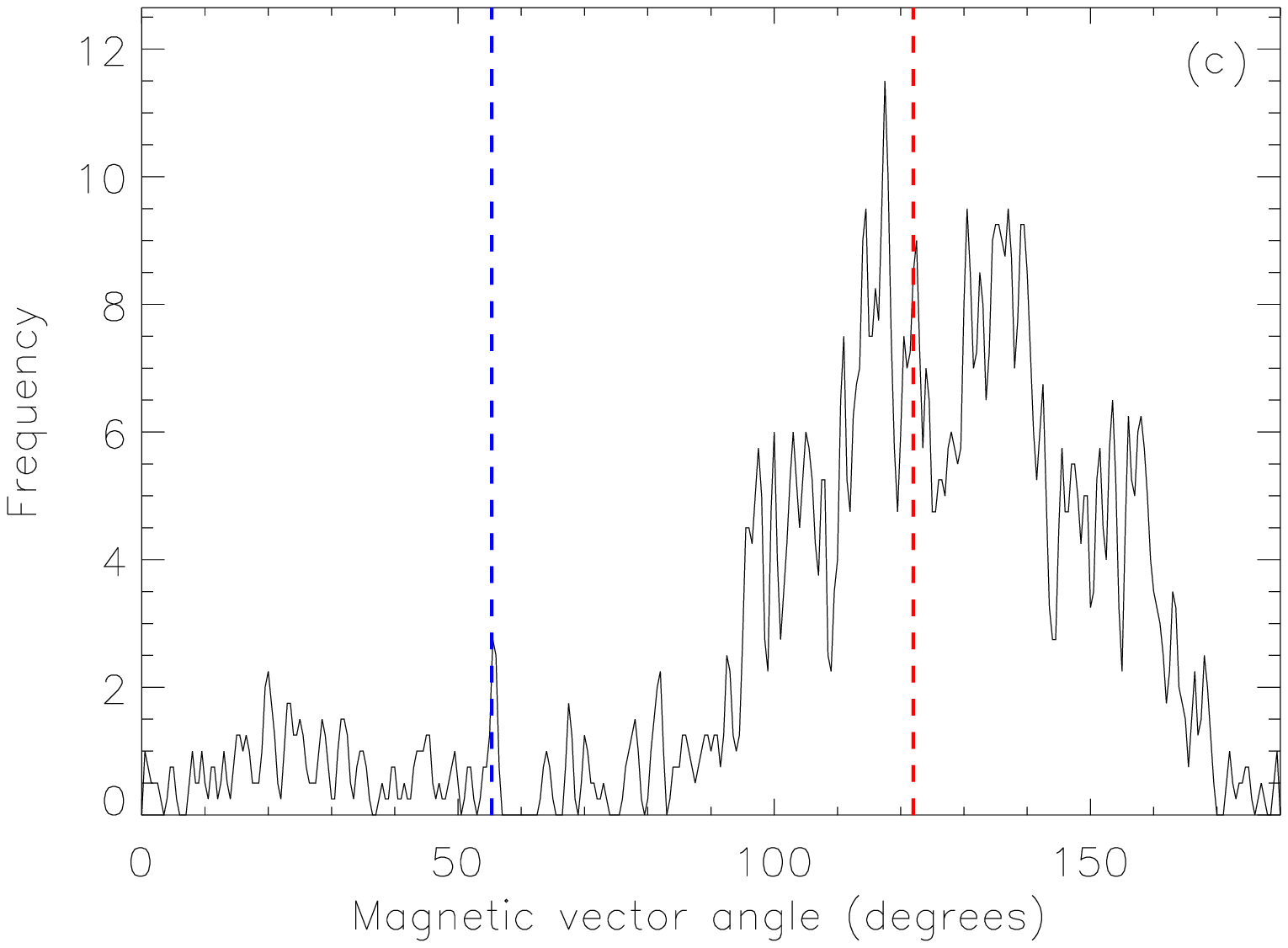}
      \hfill
  \includegraphics[width=0.475\textwidth]{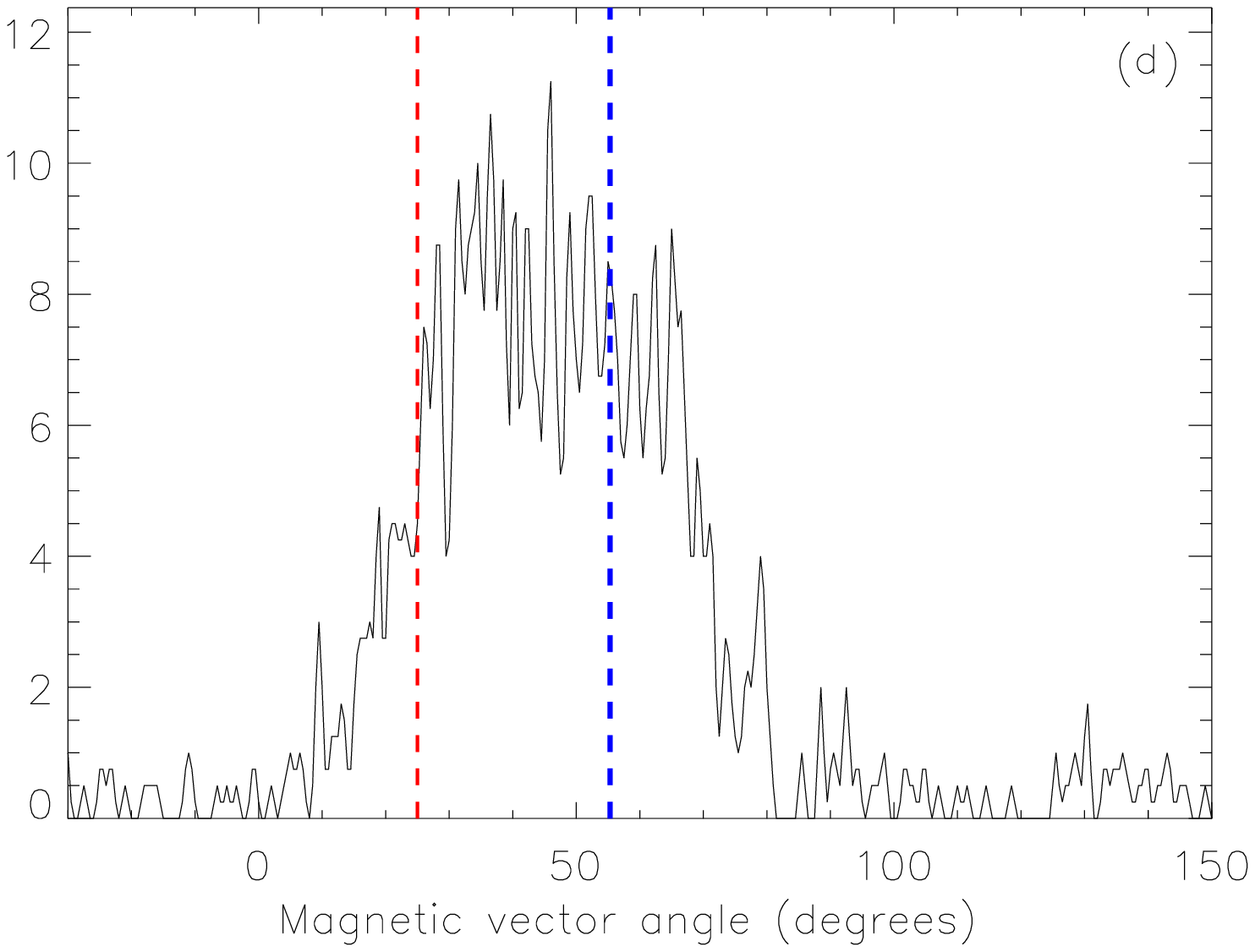}
    \hfill
    \caption{Distribution of the magnetic field vectors within the boxes indicated in Fig. \ref{vecs}. The labels at the top, right corners correspond to the respective boxes. The angles are measured counter-clockwise with respect to the north. In all frames, the perpendicular to the Galactic plane, at $55\hbox{$.\!\!{}^\circ$}3$, is plotted as a vertical dashed, blue line. In
frames (c) and (d), additional red, dashed lines represent the normal to the shock front in each case, assuming that the shock front is parallel to the flat edge of the shell.}
\label{vbox}
\end{figure*}

To illustrate the procedure described above, fits towards two random locations,
at (1) RA(2000)=8$^{\mathrm h}$23$^{\mathrm m}$11\hbox{$.\!\!{}^{\rm s}$}50, Dec.(2000)=$-43^\circ$31\hbox{$^{\prime}$}38\hbox{$.\!\!{}^{\prime\prime}$}5,
and (2) RA(2000)=8$^{\mathrm h}$22$^{\mathrm m}$53\hbox{$.\!\!{}^{\rm s}$}41, Dec.(2000)=$-42^\circ$38\hbox{$^{\prime}$}38\hbox{$.\!\!{}^{\prime\prime}$}3, 
are shown in Fig. \ref{rm-lines}. The three sets of $\chi(\lambda) - \lambda^2$ 
pairs are identified with different colors: the red line was fit with the 
frequencies numbered as 1, 5, 6, 8, and 9, the green line with 2, 3, 4, 5, and 
10, and the blue line with 1, 4, 7, 8, and 10. The corresponding points used in
the fit are represented with the same colors as the line. Some of the points 
have been shared by two sets; in such cases, they are plotted using additive 
colors, i.e., magenta for points 1 and 8 (red and blue), yellow for point 5
(red and green) and cyan for points 4 and 10 (green and blue). For location 
(1), the fits for each set of points yield RM = 2.93, 3.56, and 2.12 rad 
m$^{-2}$ respectively, resulting in an average value of $\langle{\rm RM}\rangle =
2.87\pm$0.65 rad m$^{-2}$, while for location (2), the fits give RM = 120.23,
120.42, and 120.30 rad m$^{-2}$ respectively, with an average $\langle{\rm 
RM}\rangle=120.3\pm$0.2 rad m$^{-2}$. Hence, the RM distribution shown in Fig. 
\ref{rm-map} can be considered as highly reliable.

Having solved for the $n\pi$ ambiguity resulted in a remarkable difference between the RM obtained here as compared to previous studies. Rather than the negative values (between -100 and -300 rad m$^{-2}$) inferred by \citet{MSH93} based on only two frequencies, we obtained  positive values (see Fig. \ref{rm-map}). To analyze the RM in more detail, we constructed a histogram with the statistical distribution of this parameter over Puppis A with a bin of 5 rad m$^{-2}$. The result is illustrated in Fig. \ref{rm-distr}. The RM range spans roughly from 0 to 120 rad m$^{-2}$, with a secondary peak at $\sim 170$ rad m$^{-2}$, which corresponds mostly to the yellow spot near point 1 in Fig. \ref{rm-map} discussed above. 

\subsection{Magnetic vectors}\label{mgvc}

As mentioned above, the $\chi_0$ vectors were obtained for each of the 
three groups of $\chi(\lambda) - \lambda^2$ pairs as a secondary product of 
the RM computation. Following the same method as with RM, an average map 
was obtained together with an error map. We verified that 95\% of the pixels had uncertainties below $6^\circ
\!\!.8$, so all pixels with errors exceeding this value were blanked. For the locations considered above, the values obtained were $\chi_0 = 83^\circ\!\!.16, 83^\circ\!\!.00$ and $86^\circ\!\!.01$ for
each fit at position (1), and $\chi_0 = -55^\circ\!\!.95, -55^\circ\!\!.73$ 
and $-55^\circ\!\!.40$ at position (2), with averages of $84^\circ \pm 1^\circ$
and $-55^\circ\!\!.7 \pm 0^\circ\!\!.2$ respectively. To turn electric into
magnetic vectors, a $\pi /2$ rotation was applied to $\chi_0$. The resulting spatial distribution of magnetic vectors towards Puppis A is presented in Fig. \ref{vecs}. Only one every 3 pixels in both axis is plotted, and the length of the vectors is proportional to the polarized flux at 2.24 GHz (Fig. \ref{ipol}) in such way that a 1$^{\prime}$-long vector is equivalent to 4 mJy beam$^{-1}$. To make visualization easier, magnetic vectors corresponding to polarized fluxes above 7.5 mJy beam$^{-1}$ are plotted in orange and with a different scale, in which a 1-arcmin length represents 10 mJy beam$^{-1}$.

Fig. \ref{vall} displays the statistical distribution of the magnetic angles over the whole extension of Puppis A, discarding those pixels where the flux density is below 2.5 mJy beam$^{-1}$. The distribution is clearly bi-modal, with two peaks curiously separated by about 90 degrees. These two main directions are indicated at the top, right corner of Fig. \ref{vecs}. There appears a third, narrow peak that can be represented by a Gaussian centered at $54\hbox{$.\!\!{}^\circ$}2$, with a FWHM=$7\hbox{$.\!\!{}^\circ$}7$, which is coincident with the direction normal to the Galactic plane, $55\hbox{$.\!\!{}^\circ$}3$, indicated with a vertical dashed line. Nearly 10 percent of the  magnetic vectors are contained within this peak. The green line represents a three-Gaussian fit, with peaks at $42^\circ$, $54\hbox{$.\!\!{}^\circ$}2$ and $127\hbox{$.\!\!{}^\circ$}3$. This model does not account for an excess of vectors with directions in the range $150^\circ - 170^\circ$. We will return to this issue later in Section \ref{rolvel}.

Since the earliest polarization studies toward SNRs, it became clear that young remnants tend to have radially oriented magnetic fields \citep[e.g.][]{dm76}. This is generally explained as due to Rayleigh-Taylor (R-T) fingers where magnetic fields are amplified and aligned in this direction \citep[e.g.][]{junorman}. \citet{MSH93} noted that the magnetic field in Puppis A appeared to be radial over most of the shell, consistent with its relatively young age of  3,700 yr \citep{Wink+1988}, although a more recent determination rises the age of the SNR to 5,200 yr \citep{Becker+2012}. To quantitatively explore this contention, we express all magnetic vectors in terms of a polar reference system centered at the expansion center given by \citet{Becker+2012}: RA(2000)=8$^{\mathrm h}$22$^{\mathrm m}$27\hbox{$.\!\!{}^{\rm s}$}5, Dec.(2000)=$-42^\circ$57\hbox{$^{\prime}$}29$^{\prime\prime}$. In this system, radial magnetic vectors have polar angles equal to zero. The distribution of polar angles over Puppis A is displayed in Fig. \ref{pola}. Although the peak is clearly near 0$^\circ$, the radial component is not dominant. We verified that radial vectors are prevalent on the NE and SE regions of the shell. On the other hand, a tangential behavior, typical of older SNRs, can be safely discarded.

Puppis A, however, is very peculiar in the sense that rather than circular, it depicts a  somewhat boxlike morphology. Hence, a polar reference system may not be adequate to describe a 'radial' magnetic field, which actually should be orthogonal to the shock front. Therefore, we have selected a few boxes where to scrutinize whether this direction is predominant. The selection aimed at those areas over the shell where Fig. \ref{vecs} suggests that this hypothesis could be confirmed. The boxes, parallel to the  Galactic coordinates $l,b$, are plotted in Fig. \ref{vecs}, and are labeled from {\it a} to {\it d} counter-clockwise. We include the `tail' (box {\it b}) to further analyze whether this feature is connected to Puppis A. Assuming that the orientation of the straight borders of the rim coincide in each case with the shock front, we have plotted protruding black lines for boxes {\it c} and {\it d} following the presumed shock direction (i.e., normal to the rim). In the case of box {\it a}, such direction is coincident with the perpendicular to the Galactic plane. The distribution of the magnetic vector angles in each box is displayed in Fig. \ref{vbox}. The normal to the Galactic plane is overlaid in all four frames as a dashed, blue line.

\subsection{Depolarization}\label{dep}

\begin{figure}
\centering
\includegraphics[scale=0.55]{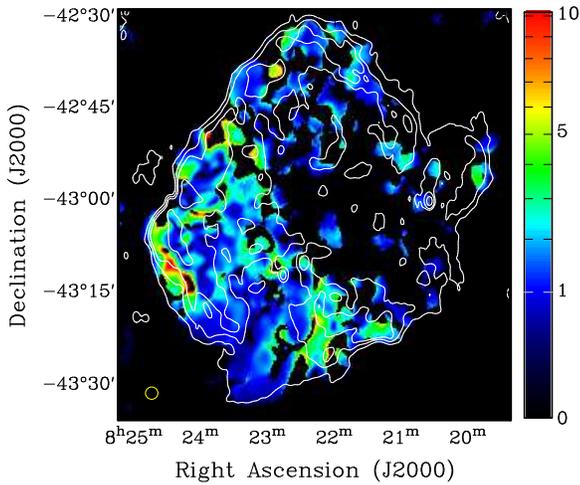}
   \caption{Depolarization ratio (2.24/1.5 GHz) over Puppis A. The color scale is displayed at the right bar. A few radio continuum white contours are overlaid to indicate the size of Puppis A. The beam is 2$^\prime \times 2^\prime$. Note that for visualization purposes, a non linear color scale was used.}
   \label{depol-map}
\end{figure}

\begin{figure}
\centering
\includegraphics[scale=0.5]{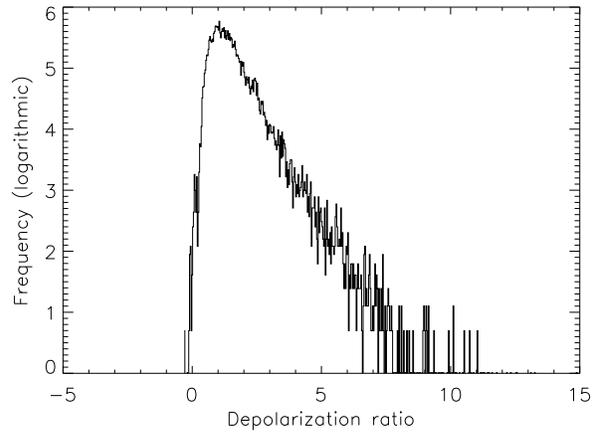}
   \caption{Logarithmic distribution of the depolarization ratio values in Puppis A. The data were binned in steps of 0.03. The highest frequency occurs at 1, indicating very low depolarization.}
     \label{depol-hist}
\end{figure}

In general, due to multiple effects like beam or bandwidth depolarization and differential Faraday rotation \citep{mch92}, the polarization rate decreases at lower frequencies. To quantify this effect, \citet{dm76} defined a `depolarization ratio' as the ratio between the polarization fraction at two frequencies, with the higher and lower frequencies as numerator and denominator respectively. We computed the depolarization ratio combining two images of the fractional polarization at 2.24 and 1.5 GHz, both convolved with a 2-arcmin circular beam. The first image was constructed as described at the end of section \ref{Obs}, while the second one was built out of {\it uv}-bands 1302 to 1686 MHz and further blanked following the same criteria as for the image at 2.24 GHz. The distribution of the depolarization ratio over Puppis A is presented in Fig. \ref{depol-map}, while Fig. \ref{depol-hist} displays a histogram of the frequency with which depolarization ratio values occur, grouped in bins of 0.03. 

Fig. \ref{depol-hist} shows that the depolarization  clearly peaks at unity, in coincidence with the result obtained by \citet{MSH93}. However, Fig. \ref{depol-map} does not reproduce the uniformity in the spatial distribution of this parameter that the authors found. Towards the southern region of Puppis A, where the RM is closer to zero (cf. Fig. \ref{rm-map}), the depolarization is closer to 1. While this coincidence is predictable according to Eq. \ref{eq:lamd}, \citet{mch92} observed that in the SNR G5.4-1.2, high depolarization correlates well with abrupt changes in RM rather than with the RM itself. As an example, we examine the feature to the NE of the shell seen in yellow in Fig. \ref{rm-map}, which has a higher RM ($\simeq 170$ rad m$^{-2}$) than the rest of the shell. At 1.5 GHz, the feature should be `Faraday-thick' \citep{BdB05,k+17}\footnote{A `Faraday-thick' medium is such that any linearly polarized emission cannot remain polarized as it passes through it, a condition usually expressed as RM$\, \lambda^2 >> 1$.}, with the RM rotating the polarization angle by almost 400$^\circ$. Besides, the Faraday effect would rotate the electric vectors by 285$^\circ$ within the spectral band used to construct the image at 1.5 GHz. However, Fig. \ref{depol-map} shows that the depolarization at this region is close to 1, meaning that the polarization observed at 2.24 GHz is not reduced when the frequency decreases to 1.5 GHz. 

\section{Discussion}\label{Disc}

\subsection{The role of the Vela SNR}\label{rolvel}

\begin{figure}
\centering
\includegraphics[scale=0.5]{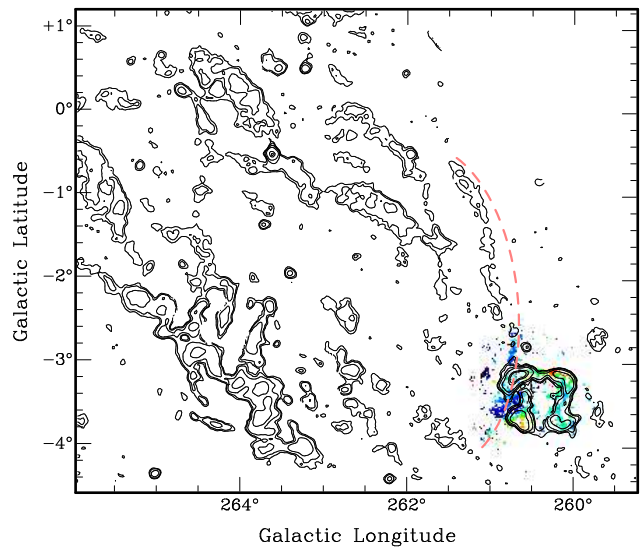}
   \caption{Distribution of the RM in Puppis A overlaid on radio continuum contours at 4.8 GHz obtained from the PMN survey. A reddish, dashed arc is included to indicate a possible correspondence between the `tail' and the outer edge of the Vela SNR, part of which is observed as the bright radio filaments in this larger field. The PMN data have a resolution of 5$^\prime$.}
     \label{vela}
\end{figure}

In Fig. \ref{rm-map}, we noticed the presence of a filamentary feature lying outside of the SNR shell, stretching out from the SE flank. There is a striking resemblance between the RM of this feature (hereafter {\it 'the tail'}) and the southern region of Puppis A, both marked by magenta boxes in Fig. \ref{rm-map}. This accordance is more evident in Fig. \ref{rm-distr}, where the RM distribution within the southern box is plotted in red and the `tail' is represented by the blue histogram. This diagram also reveals that RMs in the range 0 -- 40 rad m$^{-2}$ arise almost exclusively from the south, while the rest of Puppis A has higher RMs, mostly $\sim 75$ rad m$^{-2}$.

We observe several remarkable peculiarities that lead us to regard the `tail' as a real feature rather than an artifact resulting from the image manipulation. It is strikingly perpendicular to the Galactic plane, most magnetic vectors are aligned around this direction (note, however, another component peaking at $21\hbox{$.\!\!{}^\circ$}5$ in the distribution plot; Fig. \ref{vbox}b) and the RM is quite uniform (Fig. \ref{rm-map}) over an area of several beams. The fact that this feature has no counterpart in total intensity is not surprising, since Galactic surveys reveal several isolated bright polarization features not associated with radio continuum sources \citep[e.g.][]{dsp81,dBB05}. 

Puppis A is located in a very complex region of the sky, at the edge of the foreground Vela SNR and projected onto the Gum nebula. In Fig. \ref{vela} we overlaid the RM map with radio continuum observations at 4.85 GHz obtained from the Parkes-MIT-NRAO Radio Surveys \citep[PMN;][]{PMN-93}. The area enclosed in this figure does not fully cover the Vela SNR but suffices to reveal that the `tail' appears to be the continuation of its outermost filament. A dashed, reddish line was drawn to indicate this trend. The 2.45 GHz map of \citet{Duncan+96}, which has a better sensitivity to diffuse emission than the PMN survey, shows that the filament following the `tail' coincides with the outer edge of the Vela SNR shell. A VLBI parallax measurement of the Vela pulsar \citep{Dod+03} places this remnant at 287$^{+19}_{-17}$ pc.

The southern region of Puppis A (partly covered by the magenta box in Fig. \ref{rm-map}) is remarkably different than the rest of the SNR:  inspecting Figs. \ref{fpol}, \ref{rm-map} and \ref{depol-map}, it has a polarization fraction almost one order of magnitude higher than the rest, a lower RM (see also Fig. \ref{rm-distr}) and the lowest depolarization ($\sim 1$) throughout the full region. Based on all this evidence and the similarity with the RM at the `tail', we propose that this region is part of the Vela SNR, seen projected on Puppis A by a chance alignment in the line of sight. The low RM is compatible with the much closer distance (see Eq. \ref{eq:rm}).

With the hypothesis above in mind, we re-analyze the orientation of the magnetic vectors of Puppis A removing those coming from the southern region. The result is displayed in Fig. \ref{filt}. Interestingly, the excess between $150^\circ$ and $170^\circ$ vanishes, while the bi-modal behavior and the peak in the direction perpendicular to the Galactic plane are reinforced. At the same time, the direction of the magnetic vectors in the `tail' are tangential from the viewpoint of the Vela SNR, in agreement with the old age of this remnant \citep[11,000 yr according to the pulsar characteristic age;][]{Shan+16}. In summary, the full picture can be consistently explained considering that the southern region of Puppis A is severely contaminated by emission from the Vela SNR.

\subsection{Magnetic fields in Puppis A}\label{B-Pup}

In this section, we first estimate the line of sight component of the galactic magnetic field in the direction of Puppis A using the RM obtained. Eq. \ref{eq:rm} can be expressed in terms of the dispersion measure, DM, as \citep[e.g.][]{kothes03}
\begin{equation} \label{eq:dm} \hskip 1 cm
{\rm RM}= 0.812 B_{\parallel} {\rm DM},
\end{equation} 
where the magnetic field is given in $\mu$G and DM in pc cm$^{-3}$. DM depends both on the electron density and the distance to the source. To estimate DM for Puppis A, the Galactic electron density is required. To obtain this parameter, we run the program ymw16\footnote{The program ymw16 is available on-line at http://www.atnf.csiro.au/research/pulsar/ymw16.}, which implements the model developed by \citet{ymw16}. Searching the ATNF Pulsar Catalogue \citep{pulcat}, we found that the pulsar closest to Puppis A is PSR J0821-4221, at a $0\hbox{$.\!\!{}^\circ$}648$ projected distance from the center of the SNR shell. For this pulsar, DM=270.6 pc cm$^{-3}$ \citep{J0821}, which produces a distance of 5.8 kpc and, hence, an electron density of $n_e = 0.0107$ cm$^{-3}$. We checked this result by applying the same model to the second closest pulsar, PSR B0818-41, at a $1\hbox{$.\!\!{}^\circ$}79$ distance from the center of Puppis A. For PSR B0818-41, DM=113.4 pc cm$^{-3}$ \citep{B0818}, implying a distance of 571 pc and $n_e = 0.0107$ cm$^{-3}$, which confirms the validity of the density obtained in this region of the Galaxy. Using a distance of 1.3 $\pm 0.3$ kpc to Puppis A \citep{pap2}, we obtain DM=$144 \pm 5$ pc cm$^{-3}$. Replacing this value into Eq. \ref{eq:dm}, the component of the galactic magnetic field parallel to the line of sight turns out to be

\begin{equation} \label{eq:Bp} \hskip 1 cm
B_{\parallel} = {{\rm RM} \over {117}} \mu{\rm G}.
\end{equation}
According to the RM distribution (Figs. \ref{rm-map} and \ref{rm-distr}), and excluding the southern region which most probably belongs to the Vela SNR,
$B_{\parallel}$ varies roughly within the range  0.65 -- 1.35 $\mu$G.

In the opposite case, that is, when the foreground contribution to the Faraday effect is considered negligible and the rotation has a purely internal origin, eq. \ref{eq:rm} yields the maximum possible value for the parallel component of the magnetic field in Puppis A. Assuming that the active region has a depth of 10 pc (i.e., approximately the radius of Puppis A) and replacing $n_e$ by 1 cm$^{-3}$ \citep{Petre+82}, an upper limit of $\sim 20 \mu$G for $B_{\parallel}$ is obtained.

\begin{figure}
\centering
\includegraphics[scale=0.5]{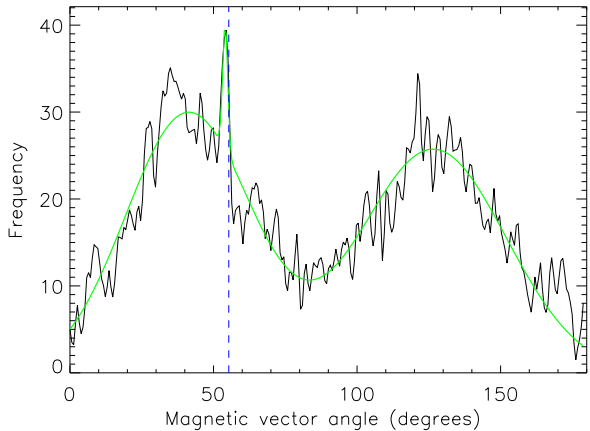}
   \caption{Same as Fig. \ref{vall} but excluding those pixels for which RM$\leq 40$ rad m$^{-2}$.}
     \label{filt}
\end{figure}

We also analyze the distribution of the direction of the magnetic vectors  in search for possible evidences of the magnetic field influence on the peculiar morphology of Puppis A. Fig. \ref{filt} provides a clue in that respect, since the two main directions in the bi-modal distribution are almost orthogonal to each other and roughly coincident with the orientation of the box-like shape depicted by Puppis A (see the upper cross in Fig. \ref{vecs}). The third, narrow peak at the direction perpendicular to the Galactic plane is striking. In SN 1006 \citep{EMR+13}, a similar peak but parallel to the Galactic plane was interpreted as an indication of the ambient magnetic field, and the two bright synchrotron lobes were then explained as polar caps in a scenario where particle acceleration was more efficient in quasi-parallel shocks, as was  later confirmed  theoretically by \citet{schneiter2015} 
and \citet{velazquez2017}. We suggest that the ISM magnetic field in Puppis A is perpendicular to the Galactic plane. Residual components of this field are probably found in box {\it a} (Fig. \ref{vbox}), while in boxes {\it c} and {\it d} the trend is not conclusive.  

On the other hand, we note a moderate evidence that the magnetic vectors are aligned with the shock direction along the shell. In box {\it a}, where the shock is parallel to the Galactic plane, there is a clear peak in the perpendicular direction, while the rest of the vectors are concentrated within $\sim 22^\circ$ around $43^\circ$.  Fig. \ref{vbox} shows that boxes {\it c} and {\it d} are marginally compatible with an alignment with the shock front direction (red, dashed lines), considering the uncertainties in the determination of the shock tilt. The error margin of the magnetic vectors, $\sim 7^\circ$, is not significant in this analysis. 

\begin{figure}
\centering
\includegraphics[scale=0.5]{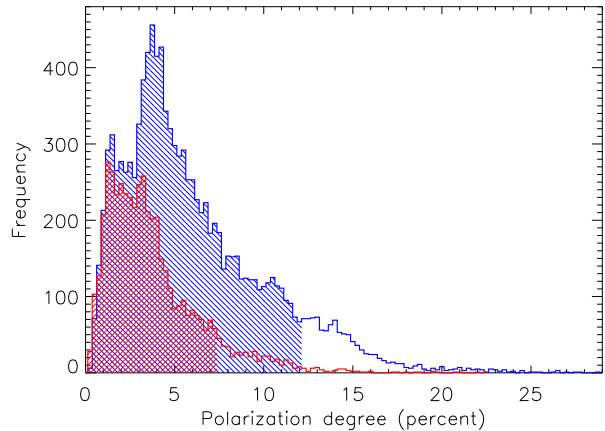}
   \caption{Statistical distribution of the fractional polarization over Puppis A based on the map in Fig. \ref{fpol}, taking all valid pixels  (blue) and filtering out those probably arising in the Vela SNR (red), using the RM$< 40$ rad m$^{-2}$ criterion as in Fig. \ref{filt}. Shaded areas contain 90\% of the pixels in each case, ammounting to 10,100 and 4,430 pixels respectively.}
     \label{FPd}
\end{figure}

\subsection{A previous wind-bubble scenario}\label{WBsc}

In Sect. \ref{rolvel}, we highlighted that the polarization degree is not uniformly distributed over Puppis A, and the highest rates are found to the south, where emission from the Vela SNR can be significant. To quantify this, we plot the statistical distribution of the polarization degree in Fig. \ref{FPd}, where the histogram is displayed in blue and the shaded area covers 90 percent of the pixels. Within this area, which excludes polarization degrees higher than 12\%, the average is $5.15 \pm 0.55$ percent, with a clear peak at $\sim 3.5$\%. To allow for confusion introduced by  the Vela SNR, we overlaid a second plot (in red) filtering out all pixels where RM$< 40$ rad m$^{-2}$. Again, the shaded area encloses 90\% of the valid pixels, leaving outside polarization degrees higher than 7\%, and lowering the average to  $3.1 \pm 0.5$ percent, with peaks at 1.3\% and 3\%. It is clear that the region with the highest polarization comes from the Vela SNR.

Our study confirms a very low polarization over most of Puppis A, as measured by \citet{md75} at 5 GHz and by \citet{MSH93} at 4.75 GHz and 8.4 GHz. The synchrotron emission theory predicts that an optically thin source in a uniform magnetic field will emit polarized radiation with a polarization fraction of  $p \leq (\alpha +1)/(\alpha +5/3)$, where $\alpha$ is the spectral index defined as $S_\nu \propto \nu^{-\alpha}$ \citep[e.g.][]{pach73}. Since $\alpha= 0.563$ for Puppis A \citep{pap1}, the highest theoretical fractional polarization is $p_{max} \simeq 70$ percent. The ratio between the observed fractional polarization and $p_{max}$ is related to the ratio between the magnetic energies of the uniform field component and the total field \citep{burn66}: $p/p_{max} \simeq E_{\rm B_{\rm uniform}} / E_{\rm B_{\rm total}}$. Hence, only a few percent (in the range $\sim 1.8$ to 4.5) of the magnetic field is ordered in Puppis A.

As mentioned in Sect. \ref{dep}, several reasons reduce the theoretical  polarization degrees to the observed levels (beam or bandwidth depolarization, differential Faraday rotation, etc). However, Puppis A appears among the SNRs with the lowest fractional polarizations observed \citep[e.g.][]{rg93,CGPS-SNRs,sino6l}. Such depolarization can occur if the SNR shock front evolves within a very turbulent magnetized medium. \citet{spohl09} find that an expansion into a strong, isotropic turbulence with a flat spectrum and with no shock compression can result in very low polarization degrees ($\lesssim 5$ percent). This scenario is also consistent with the model developed by \citet{BP16}. The low linear polarization observed in Puppis A requires that the random component of the magnetic field, $\mathbf{\delta B}$, in the pre-shock medium be twice the ordered magnetic field or even higher. 

The origin of the pre-shock turbulent magnetized medium can be explained invoking the scenario proposed by \citet{pap2}, in which Puppis A is contained in the stellar bubble produced by its progenitor during its O-type phase. Taking into account that after the main sequence (MS) phase,  massive stars go through other evolutionary phases,  in which they have different types of winds and generate a variety of structures, it is expected that the medium where the explosion took place was not uniform. 
Stars having an initial mass in the range 25 -- 40 M$_{\odot}$, after leaving the MS enter the red supergiant (RSG) or luminous blue variable (LBV) phase, exhibiting dense, slow, and dust-rich winds that expand into the ISM. This slow wind expels more than half of the initial mass of the star, exposing its hot core and becoming a Wolf-Rayet (WR) star. Then, the strong winds of the WR star sweep up and compress the previously ejected RSG or LBV material into a shell creating the so-called  WR nebulae \citep{cro07}. 

\citet{dwarkadas2007} analyzed the evolution of  supernovae in circumstellar wind bubbles using multidimensional simulations. His study revealed the presence of density and pressure fluctuations within the surrounding medium, as well as of hydrodynamic instabilities, the growth of vorticity, and the onset of turbulence.
\citet{dwarkadas2007} showed that the interaction between the RSG and WR winds provides the appropriate conditions for the R-T instability to develop, producing turbulence within the interior of the MS bubble (see his Fig. 16). Considering that this scenario is probable for the progenitor star of Puppis A, the expansion could be taking place into a turbulent medium. We can assume that any pre-existing IS magnetic field was frozen with the turbulent gas. Hence, the SNR shock front would evolve into an equally disorganized magnetic field, resulting in the low linear polarization observed. 

An expansion inside the stellar bubble blown by the progenitor star may have an extra consequence: the unique shape of Puppis A. \citet{garciasegura2000} carried out several 3-dimensional magneto-hydrodynamical (MHD) simulations in order to explain the bipolar and elliptical morphologies observed in planetary nebulae and found that the models needed to consider interacting stellar winds, one of them magnetized. When the axis of the magnetic dipole is tilted with respect to the stellar rotation axis and the wind outflow is elliptical rather than bipolar (i.e., the equatorial component is non-negligible), the authors reproduce a box-like morphology similar to that of Puppis A (see their Fig. 6a, corresponding to model E3). In their model, the stellar wind magnetic field has a negligible radial component B$_{\mathrm r}$ \citep[since B$_{\mathrm r}$ decreases with the distance $\mathrm r$ to the central star as $1/{\mathrm r}^2$;][]{garcias1999} and an azimuthal component B$_{\phi}$ given by: 
\begin{equation}
\mathrm{B_{\phi}=B_* \frac{v_{rot}}{v_{\infty}}
\frac{R_*}{\mathrm r}\ sin\ \theta},
\label{bphi}
\end{equation}
 where B$_*$ is the magnetic field strength at the stellar surface, $\mathrm{R_*}$ is the RSG radius, $\mathrm{v_{rot}}$ is the rotational velocity, $\mathrm{v_{\infty}}$ is the terminal wind velocity and $\theta$ is the polar angle.  Considering $\mathrm{B_*=2~G}$, $\mathrm{v_{rot}=5.6\ km\ s^{-1}}$, $\mathrm{v_{\infty}=100\ km\ s^{-1}}$, $\mathrm{R_*=10\ R_{\odot}}$ \citep{garciasegura2000,garcias1999},   $\mathrm{B_{\phi}}$ decreases to $\mathrm{2\times 10^{-7}}$ G for $\mathrm{r}=0.1$~pc, the typical size of the nebula corresponding to their model E3. This model is characterized by a ratio between magnetic energy density and kinetic energy density, $\sigma$, of 0.01. The parameter $\sigma$ is important in determining the final morphology of the planetary nebula. Although the progenitors of planetary nebulae are low-mass stars, the underlying models coincide in that both involve stellar bubbles produced by interacting winds, and the main difference are the parameters considered for the stellar wind of the SN progenitor \citep[see for example][]{udDoula2017}. For the case of the Puppis A progenitor star, we can assume $\mathrm{B_*} = 300$ G for a stellar RSG radius of 290 $\mathrm{R_{\odot}}$, a ratio $\mathrm{v_{rot}/v_{\infty}}$ of 0.1 \citep{moranchel17} and, by using Eq. (\ref{bphi}), we estimate  $\mathrm{B_{\phi}} = 2\times 10^{-5}$ G for $\mathrm{r=10\ pc}$ (the SNR radius). With these values for the RSG stellar wind, we estimated the ratio between the magnetic energy density of the stellar wind bubble and the SNR energy density (assumed to be $5\times 10^{50}$~erg uniformly distributed throughout the whole SNR volume), yielding $\sigma \simeq 0.02$, which is quite similar to the value obtained for the planetary nebula of model E3 \citep{garciasegura2000}. Then, the final morphology of Puppis A is expected to be the result of the whole evolutionary story inside a magnetized stellar wind bubble, i.e. after the star explodes as a SN, the SNR shock wave will lose its initially spherical morphology, and may acquire an elliptical or a box-like morphology.

As mentioned above, the gas within the stellar bubble is turbulent, which favors the development of R-T instabilities at the contact discontinuity, behind the SNR shock front. In a spherical expansion, the R-T fingers are radially oriented, which could explain the radial component of the magnetic field observed in young remnants. Likewise, a box-like (or elliptical) expansion would drive the R-T fingers to grow orthogonal to the shock front. We then suggest that the two main directions of the magnetic field observed in Fig. \ref{filt} can be the combined result of a box-like SNR expansion and the development of R-T instabilities. A detailed analysis of this hypothesis is beyond the scope of the present paper and will be explored in a future work.

\section{Conclusions}\label{Conc}

We present an analysis of the polarization emission towards the SNR Puppis A with unprecedented detail. The capabilities of the Compact Array Broadband Backend (CABB) of the ATCA made it possible to obtain polarization maps at several nearby frequencies by splitting its 2 GHz wideband into a number of 128 MHz-wide bands. This technique allowed us to solve for the $n\pi$ indetermination and therefore accurately compute the RM and retrieve the original direction of the polarization vectors in the SNR. 

The high quality RM map obtained revealed that the southern region of Puppis A is mostly foreground emission from the Vela SNR, incidentally seen in projection. We gathered substantial evidence to confirm this result, like low RM and depolarization in this region, together with polarization degrees and magnetic vector directions not fully compatible with the rest of Puppis A. When the contribution from the Vela SNR is filtered out, the statistical behavior of the magnetic vectors clearly shows two almost orthogonal preferential directions, which roughly follow the peculiar box-like morphology of this SNR. A third, narrow peak at a direction perpendicular to the Galactic plane is probably a residual indication of a uniform IS magnetic field in this region of the Galaxy. 

Our findings provide a crucial contribution to the long lasting debate about the reason behind the morphology of Puppis A, which we propose could be determined by a stellar wind bubble driven by the SN progenitor and not by strong ISM density gradients or an interaction with molecular clouds. The stellar bubble consistent with our scenario is the result of the interaction of the different types of winds, one of them magnetized, along the life of the SN progenitor, which can produce a box-like cavity, as those obtained by \citet{garciasegura2000} for the case of planetary nebulae. On the other hand, the very low polarization degree measured across most of the remnant can be explained by the expansion of the SNR shock front into the turbulent medium of the stellar bubble, whose magnetic field has an important random component.

In summary, the very low polarization degree, the distribution of the magnetic field vectors and the box-like morphology observed for the SNR Puppis A altogether conform a consistent picture where the remnant evolves into a stellar bubble created by the interaction of the stellar winds launched by the SN progenitor during the different phases spanned along its life, provided that one of the winds is magnetized and the magnetic dipole is tilted with respect to the star rotation axis.

\section*{Acknowledgements}

We appreciate useful discussions with Laura Richter, Timothy Shimwell and Mark Wieringa for solving the mosaic field name length bug. We also acknowledge Alejandro Esquivel for a critical reading of this manuscript. This research was partially funded by CONICET grant PIP 112-201207-00226. PFV acknowledges financial support from DGAPA-UNAM grant IG120218/IG100218. The Australia Telescope Compact Array is part of the Australia Telescope National Facility which is funded by the Commonwealth of Australia for operation as a National Facility managed by CSIRO. EMR and SC are members of the Carrera del Investigador Cient\'\i fico of CONICET, Argentina.

\bibliographystyle{mn2e}
\bibliography{apj-jour,puppol}{}

\begin{thebibliography}{}

\bibitem[\protect\citeauthoryear{{Arendt}, {Dwek}, {Blair}, {Ghavamian},
  {Hwang}, {Long}, {Petre}, {Rho} \& {Winkler}}{{Arendt} et~al.}{2010}]{IR2010}
{Arendt} R.~G.,  {Dwek} E.,  {Blair} W.~P.,  {Ghavamian} P.,  {Hwang} U.,
  {Long} K.~S.,  {Petre} R.,  {Rho} J.,    {Winkler} P.~F.,  2010, \apj, 725,
  585

\bibitem[\protect\citeauthoryear{{Arthur} \& {Falle}}{{Arthur} \&
  {Falle}}{1991}]{af91}
{Arthur} S.~J.,  {Falle} S.~A.~E.~G.,  1991, \mnras, 251, 93

\bibitem[\protect\citeauthoryear{{Arzoumanian}, {Nice}, {Taylor} \&
  {Thorsett}}{{Arzoumanian} et~al.}{1994}]{B0818}
{Arzoumanian} Z.,  {Nice} D.~J.,  {Taylor} J.~H.,    {Thorsett} S.~E.,  1994,
  \apj, 422, 671

\bibitem[\protect\citeauthoryear{{Bandiera} \& {Petruk}}{{Bandiera} \&
  {Petruk}}{2016}]{BP16}
{Bandiera} R.,  {Petruk} O.,  2016, \mnras, 459, 178

\bibitem[\protect\citeauthoryear{{Becker}, {Prinz}, {Winkler} \&
  {Petre}}{{Becker} et~al.}{2012}]{Becker+2012}
{Becker} W.,  {Prinz} T.,  {Winkler} P.~F.,    {Petre} R.,  2012, \apj, 755,
  141

\bibitem[\protect\citeauthoryear{{Brentjens} \& {de Bruyn}}{{Brentjens} \& {de
  Bruyn}}{2005}]{BdB05}
{Brentjens} M.~A.,  {de Bruyn} A.~G.,  2005, \aap, 441, 1217

\bibitem[\protect\citeauthoryear{{Burgay}, {Joshi}, {D'Amico}, {Possenti},
  {Lyne}, {Manchester}, {McLaughlin}, {Kramer}, {Camilo} \& {Freire}}{{Burgay}
  et~al.}{2006}]{J0821}
{Burgay} M.,  {Joshi} B.~C.,  {D'Amico} N.,  {Possenti} A.,  {Lyne} A.~G.,
  {Manchester} R.~N.,  {McLaughlin} M.~A.,  {Kramer} M.,  {Camilo} F.,
  {Freire} P.~C.~C.,  2006, \mnras, 368, 283

\bibitem[\protect\citeauthoryear{{Burn}}{{Burn}}{1966}]{burn66}
{Burn} B.~J.,  1966, \mnras, 133, 67

\bibitem[\protect\citeauthoryear{{Crowther}}{{Crowther}}{2007}]{cro07}
{Crowther} P.~A.,  2007, \araa, 45, 177

\bibitem[\protect\citeauthoryear{{de Bruyn} \& {Brentjens}}{{de Bruyn} \&
  {Brentjens}}{2005}]{dBB05}
{de Bruyn} A.~G.,  {Brentjens} M.~A.,  2005, \aap, 441, 931

\bibitem[\protect\citeauthoryear{{Dickel} \& {Milne}}{{Dickel} \&
  {Milne}}{1976}]{dm76}
{Dickel} J.~R.,  {Milne} D.~K.,  1976, Australian Journal of Physics, 29, 435

\bibitem[\protect\citeauthoryear{{Dodson}, {Legge}, {Reynolds} \&
  {McCulloch}}{{Dodson} et~al.}{2003}]{Dod+03}
{Dodson} R.,  {Legge} D.,  {Reynolds} J.~E.,    {McCulloch} P.~M.,  2003, \apj,
  596, 1137

\bibitem[\protect\citeauthoryear{{Downes}, {Salter} \& {Pauls}}{{Downes}
  et~al.}{1981}]{dsp81}
{Downes} A.~J.~B.,  {Salter} C.~J.,    {Pauls} T.,  1981, \aap, 97, 296

\bibitem[\protect\citeauthoryear{{Dubner}, {Loiseau},
  {Rodr{\'{\i}}guez-Pascual}, {Smith}, {Giacani} \& {Castelletti}}{{Dubner}
  et~al.}{2013}]{Xgd+13}
{Dubner} G.,  {Loiseau} N.,  {Rodr{\'{\i}}guez-Pascual} P.,  {Smith} M.~J.~S.,
  {Giacani} E.,    {Castelletti} G.,  2013, \aap, 555, A9

\bibitem[\protect\citeauthoryear{{Dubner} \& {Arnal}}{{Dubner} \&
  {Arnal}}{1988}]{gd+ma88}
{Dubner} G.~M.,  {Arnal} E.~M.,  1988, \aaps, 75, 363

\bibitem[\protect\citeauthoryear{{Duncan}, {Stewart}, {Haynes} \&
  {Jones}}{{Duncan} et~al.}{1996}]{Duncan+96}
{Duncan} A.~R.,  {Stewart} R.~T.,  {Haynes} R.~F.,    {Jones} K.~L.,  1996,
  \mnras, 280, 252

\bibitem[\protect\citeauthoryear{{Dwarkadas}}{{Dwarkadas}}{2007}]{dwarkadas2007}
{Dwarkadas} V.~V.,  2007, \apj, 667, 226

\bibitem[\protect\citeauthoryear{{Frail}, {Goss}, {Reynoso}, {Giacani}, {Green}
  \& {Otrupcek}}{{Frail} et~al.}{1996}]{frail+96}
{Frail} D.~A.,  {Goss} W.~M.,  {Reynoso} E.~M.,  {Giacani} E.~B.,  {Green}
  A.~J.,    {Otrupcek} R.,  1996, \aj, 111, 1651

\bibitem[\protect\citeauthoryear{{Fulbright} \& {Reynolds}}{{Fulbright} \&
  {Reynolds}}{1990}]{fr90}
{Fulbright} M.~S.,  {Reynolds} S.~P.,  1990, \apj, 357, 591

\bibitem[\protect\citeauthoryear{{Garc{\'{\i}}a-Segura}, {Langer},
  {R{\'o}{\.z}yczka} \& {Franco}}{{Garc{\'{\i}}a-Segura}
  et~al.}{1999}]{garcias1999}
{Garc{\'{\i}}a-Segura} G.,  {Langer} N.,  {R{\'o}{\.z}yczka} M.,    {Franco}
  J.,  1999, \apj, 517, 767

\bibitem[\protect\citeauthoryear{{Garc{\'{\i}}a-Segura} \&
  {L{\'o}pez}}{{Garc{\'{\i}}a-Segura} \& {L{\'o}pez}}{2000}]{garciasegura2000}
{Garc{\'{\i}}a-Segura} G.,  {L{\'o}pez} J.~A.,  2000, \apj, 544, 336

\bibitem[\protect\citeauthoryear{{Griffith} \& {Wright}}{{Griffith} \&
  {Wright}}{1993}]{PMN-93}
{Griffith} M.~R.,  {Wright} A.~E.,  1993, \aj, 105, 1666

\bibitem[\protect\citeauthoryear{{Jun} \& {Norman}}{{Jun} \&
  {Norman}}{1996}]{junorman}
{Jun} B.-I.,  {Norman} M.~L.,  1996, \apj, 472, 245

\bibitem[\protect\citeauthoryear{{Kothes}}{{Kothes}}{2003}]{kothes03}
{Kothes} R.,  2003, \aap, 408, 187

\bibitem[\protect\citeauthoryear{{Kothes}, {Fedotov}, {Foster} \&
  {Uyan{\i}ker}}{{Kothes} et~al.}{2006}]{CGPS-SNRs}
{Kothes} R.,  {Fedotov} K.,  {Foster} T.~J.,    {Uyan{\i}ker} B.,  2006, \aap,
  457, 1081

\bibitem[\protect\citeauthoryear{{Kothes}, {Sun}, {Gaensler} \&
  {Reich}}{{Kothes} et~al.}{2018}]{k+17}
{Kothes} R.,  {Sun} X.,  {Gaensler} B.,    {Reich} W.,  2018, \apj, 852, 54

\bibitem[\protect\citeauthoryear{{Manchester}, {Hobbs}, {Teoh} \&
  {Hobbs}}{{Manchester} et~al.}{2005}]{pulcat}
{Manchester} R.~N.,  {Hobbs} G.~B.,  {Teoh} A.,    {Hobbs} M.,  2005, \aj, 129,
  1993

\bibitem[\protect\citeauthoryear{{Milne}, {Caswell} \& {Haynes}}{{Milne}
  et~al.}{1992}]{mch92}
{Milne} D.~K.,  {Caswell} J.~L.,    {Haynes} R.~F.,  1992, \mnras, 255, 707

\bibitem[\protect\citeauthoryear{{Milne} \& {Dickel}}{{Milne} \&
  {Dickel}}{1975}]{md75}
{Milne} D.~K.,  {Dickel} J.~R.,  1975, Australian Journal of Physics, 28, 209

\bibitem[\protect\citeauthoryear{{Milne}, {Stewart} \& {Haynes}}{{Milne}
  et~al.}{1993}]{MSH93}
{Milne} D.~K.,  {Stewart} R.~T.,    {Haynes} R.~F.,  1993, \mnras, 261, 366

\bibitem[\protect\citeauthoryear{{Moranchel-Basurto}, {Vel{\'a}zquez},
  {Giacani}, {Toledo-Roy}, {Schneiter}, {De Colle} \&
  {Esquivel}}{{Moranchel-Basurto} et~al.}{2017}]{moranchel17}
{Moranchel-Basurto} A.,  {Vel{\'a}zquez} P.~F.,  {Giacani} E.,  {Toledo-Roy}
  J.~C.,  {Schneiter} E.~M.,  {De Colle} F.,    {Esquivel} A.,  2017, \mnras,
  472, 2117

\bibitem[\protect\citeauthoryear{{Pacholczyk}}{{Pacholczyk}}{1973}]{pach73}
{Pacholczyk} A.~G.,  1973, {Radio astrophysics. Non-thermal processes in
  galactic and extragalactic sources.}

\bibitem[\protect\citeauthoryear{{Petre}, {Kriss}, {Winkler} \&
  {Canizares}}{{Petre} et~al.}{1982}]{Petre+82}
{Petre} R.,  {Kriss} G.~A.,  {Winkler} P.~F.,    {Canizares} C.~R.,  1982,
  \apj, 258, 22

\bibitem[\protect\citeauthoryear{{Reynolds}, {Gaensler} \&
  {Bocchino}}{{Reynolds} et~al.}{2012}]{rgb11}
{Reynolds} S.~P.,  {Gaensler} B.~M.,    {Bocchino} F.,  2012, \ssr, 166, 231

\bibitem[\protect\citeauthoryear{{Reynolds} \& {Gilmore}}{{Reynolds} \&
  {Gilmore}}{1993}]{rg93}
{Reynolds} S.~P.,  {Gilmore} D.~M.,  1993, \aj, 106, 272

\bibitem[\protect\citeauthoryear{{Reynoso}, {Cichowolski} \& {Walsh}}{{Reynoso}
  et~al.}{2017}]{pap2}
{Reynoso} E.~M.,  {Cichowolski} S.,    {Walsh} A.~J.,  2017, \mnras, 464, 3029

\bibitem[\protect\citeauthoryear{{Reynoso}, {Dubner}, {Goss} \&
  {Arnal}}{{Reynoso} et~al.}{1995}]{EMR+95}
{Reynoso} E.~M.,  {Dubner} G.~M.,  {Goss} W.~M.,    {Arnal} E.~M.,  1995, \aj,
  110, 318

\bibitem[\protect\citeauthoryear{{Reynoso}, {Hughes} \& {Moffett}}{{Reynoso}
  et~al.}{2013}]{EMR+13}
{Reynoso} E.~M.,  {Hughes} J.~P.,    {Moffett} D.~A.,  2013, \aj, 145, 104

\bibitem[\protect\citeauthoryear{{Reynoso} \& {Walsh}}{{Reynoso} \&
  {Walsh}}{2015}]{pap1}
{Reynoso} E.~M.,  {Walsh} A.~J.,  2015, \mnras, 451, 3044

\bibitem[\protect\citeauthoryear{{Sault}, {Teuben} \& {Wright}}{{Sault}
  et~al.}{1995}]{Sault+95}
{Sault} R.~J.,  {Teuben} P.~J.,    {Wright} M.~C.~H.,  1995, in {R.~A.~Shaw,
  H.~E.~Payne, \& J.~J.~E.~Hayes} ed., Astronomical Data Analysis Software and
  Systems IV Vol.~77 of Astronomical Society of the Pacific Conference Series,
  {A Retrospective View of MIRIAD}.
pp 433--+

\bibitem[\protect\citeauthoryear{{Schneiter}, {Vel{\'a}zquez}, {Reynoso},
  {Esquivel} \& {De Colle}}{{Schneiter} et~al.}{2015}]{schneiter2015}
{Schneiter} E.~M.,  {Vel{\'a}zquez} P.~F.,  {Reynoso} E.~M.,  {Esquivel} A.,
  {De Colle} F.,  2015, \mnras, 449, 88

\bibitem[\protect\citeauthoryear{{Shannon}, {Lentati}, {Kerr}, {Johnston},
  {Hobbs} \& {Manchester}}{{Shannon} et~al.}{2016}]{Shan+16}
{Shannon} R.~M.,  {Lentati} L.~T.,  {Kerr} M.,  {Johnston} S.,  {Hobbs} G.,
  {Manchester} R.~N.,  2016, \mnras, 459, 3104

\bibitem[\protect\citeauthoryear{{Stroman} \& {Pohl}}{{Stroman} \&
  {Pohl}}{2009}]{spohl09}
{Stroman} W.,  {Pohl} M.,  2009, \apj, 696, 1864

\bibitem[\protect\citeauthoryear{{Sun}, {Reich}, {Reich}, {Xiao}, {Gao} \&
  {Han}}{{Sun} et~al.}{2011}]{sino6l}
{Sun} X.~H.,  {Reich} P.,  {Reich} W.,  {Xiao} L.,  {Gao} X.~Y.,    {Han}
  J.~L.,  2011, \aap, 536, A83

\bibitem[\protect\citeauthoryear{{ud-Doula}}{{ud-Doula}}{2017}]{udDoula2017}
{ud-Doula} A.,  2017, Astronomische Nachrichten, 338, 944

\bibitem[\protect\citeauthoryear{{Vel{\'a}zquez}, {Schneiter}, {Reynoso},
  {Esquivel}, {De Colle}, {Toledo-Roy}, {G{\'o}mez}, {Sieyra} \&
  {Moranchel-Basurto}}{{Vel{\'a}zquez} et~al.}{2017}]{velazquez2017}
{Vel{\'a}zquez} P.~F.,  {Schneiter} E.~M.,  {Reynoso} E.~M.,  {Esquivel} A.,
  {De Colle} F.,  {Toledo-Roy} J.~C.,  {G{\'o}mez} D.~O.,  {Sieyra} M.~V.,
  {Moranchel-Basurto} A.,  2017, \mnras, 466, 4851

\bibitem[\protect\citeauthoryear{{Wilson}, {Ferris}, {Axtens}, {Brown},
  {Davis}, {Hampson}, {Leach}, {Roberts}, {Saunders}, {Koribalski}, {Caswell},
  {Lenc}, {Stevens}, {Voronkov}, {Wieringa} \& et al.}{{Wilson}
  et~al.}{2011}]{CABB+11}
{Wilson} W.~E.,  {Ferris} R.~H.,  {Axtens} P.,  {Brown} A.,  {Davis} E.,
  {Hampson} G.,  {Leach} M.,  {Roberts} P.,  {Saunders} S.,  {Koribalski}
  B.~S.,  {Caswell} J.~L.,  {Lenc} E.,  {Stevens} J.,  {Voronkov} M.~A.,
  {Wieringa} M.~H.,    et al. 2011, \mnras, 416, 832

\bibitem[\protect\citeauthoryear{{Winkler}, {Tuttle}, {Kirshner} \&
  {Irwin}}{{Winkler} et~al.}{1988}]{Wink+1988}
{Winkler} P.~F.,  {Tuttle} J.~H.,  {Kirshner} R.~P.,    {Irwin} M.~J.,  1988,
  in {Roger} R.~S.,  {Landecker} T.~L.,  eds, IAU Colloq. 101: Supernova
  Remnants and the Interstellar Medium {Kinematics of Oxygen-Rich Filaments in
  Puppis A}.
p.~65

\bibitem[\protect\citeauthoryear{{Woermann}, {Gaylard} \&
  {Otrupcek}}{{Woermann} et~al.}{2000}]{beate+00}
{Woermann} B.,  {Gaylard} M.~J.,    {Otrupcek} R.,  2000, \mnras, 317, 421

\bibitem[\protect\citeauthoryear{{Yao}, {Manchester} \& {Wang}}{{Yao}
  et~al.}{2017}]{ymw16}
{Yao} J.~M.,  {Manchester} R.~N.,    {Wang} N.,  2017, \apj, 835, 29

\end{thebibliography}

\bsp

\label{lastpage}

\end{document}